\documentclass{aa}
\usepackage{graphics}
\begin{document}

  \thesaurus{04           
               (10.19.3;  
                13.09.1)} 

\title{Near Infrared Observations of the Local Arm}

\author{C. Porcel \inst{1}, F. Garz\'on \inst{2}, J. Jim\'enez-Vicente \inst{1} \and E. Battaner \inst{1}}

\institute{Departamento de F{\'\i}sica Te\'orica y del
 Cosmos, Universidad de Granada, Granada, Spain
\and Instituto de Astrof{\'\i}sica de Canarias. E-38200, La Laguna. Tenerife. Spain}

\date{Received; accepted}
\maketitle

\begin{abstract}

Observations of the Local Arm have been carried out in the
near-infrared with the 1.5 m ``Car\-los S\'an\-chez'' telescope in
Tenerife. A model of the disc with adjustable parameters fitted
to reproduce the DIRBE-COBE survey, has been subtracted from the
observational data in order to obtain a clean map of the Local
Arm, uncontaminated by other components of the disc. The arm is
more than 70 pc over the plane and is wider than 200 pc. At a latitude of about $80^\circ$ the deviation of the arm from the galactic plane is so large that was only partially observed with our observational window of $\triangle b = \pm 6^\circ$. The
elevation over the plane vanishes and the width  decreases as
the Arm comes closer to the Sun.

\end{abstract}

\section{Introduction}
Because of its proximity, considerable attention has been paid
to the spiral feature known as the Local Arm. Optical and 21 cm
observations have long been available. Buss et al. (1994)
in the ultraviolet, Fatoohi et al. (1996) in gamma-rays, Oliver
et al. (1996) in CO, have, more recently, among others,
reported observations at other wavelengths.
Dynamic  properties have been analyzed by Palous (1987),
Comeron \& Torra (1991) and others.

In the near infrared, recent surveys (e.g. Odenwald \&
Schwartz 1993; Freudenreich et al. 1994; Garz\'on et al. 1993)
have provided significant information, but the application of
these to the study of the Local Arm has been insufficient. Ortiz \&
Lepine (1993) themselves found  their description
of the Local Arm to be  unsatisfactory. More infrared 
observations are necessary and
the main purpose of this work is to compensate for this shortcoming.

We report our own observations with the 1.5 ``Carlos S\'anchez'' telescope in
Tenerife, used to search the infrared morphology of the closest
spiral feature; this was undertaken bearing in mind the
following goals:

1.- To observe the first quadrant region of the arm, which in
practice limits  attention to the longitude range
$70^\circ-90^\circ$. Observations extend to slightly higher
longitudes, but here  interpretation becomes difficult
because the arm is too close.

2.- To use high spatial resolution techniques in order to
observe individual stars. We determine stellar counts instead of
total fluxes. We can in this way perform detailed analysis at various magnitude ranges. The immediate advantage is that this  procedure
constitutes a natural filter: very bright magnitude stars must,
necessarily, be nearby. The importance of the contribution
 of other components
is thus minimized. This probably compensates for the
considerable observation time, which in practice means a reduced
 number of scans. COBE data, for instance, are poorer resolution  flux data, and are  therefore less suitable to study a close feature like the Local Arm, as they are contaminated by the contribution of larger scale features behind it.

3.- To discriminate between the different contributions to the star counts
from other galactic components, mainly the exponential component
of the disc. It is to be emphasized that the northern warp of
our galaxy is just behind the Local Arm in its first quadrant portion.

\section{Observations}
Observations were made as part of the TMGS (`` Two Micron
Galactic Survey'') (Garz\'on et al. 1993; Hammersley et al.
1994; Calbet et al. 1995) carried out by the IAC (Instituto de
Astrof{\'\i}sica de Canarias).
The aim is the survey in the K-band of the galactic plane with a
15 arcsec resolution and a 10.5 limit magnitude (Calbet, 1993);
therefore, individual stars are observed, so that position and
magnitude are determined for each star.
Before the observations made for this work, the survey was
mainly devoted to the galactic centre and the order of scans was
altered to observe the 
$70^\circ-100^\circ$ galactic longitude zone, where  the
Local Arm lies. The original purpose was the study of the north
warp in the near infrared, but in this zone the emission is
dominated by the Local Arm. The obtained data were therefore
used to describe the NIR Local Arm.

For TMGS the 1.5 m NIR  ``Carlos S\'anchez''  telescope is used.
The telescope is located in Tenerife at 2400 m altitude. Scans
are obtained by switching off the right ascension telescope
motor, and thus constant declination scans are obtained, taking
advantage of  terrestrial rotation. We do not further describe the
observation techniques and the  system  for data acquisition and
processing as this has been extensively done on previous occasions
 (Hammersley  1989; Garz\'on et al. 1993)

\begin{figure}
\resizebox{8.8cm}{!}{\includegraphics{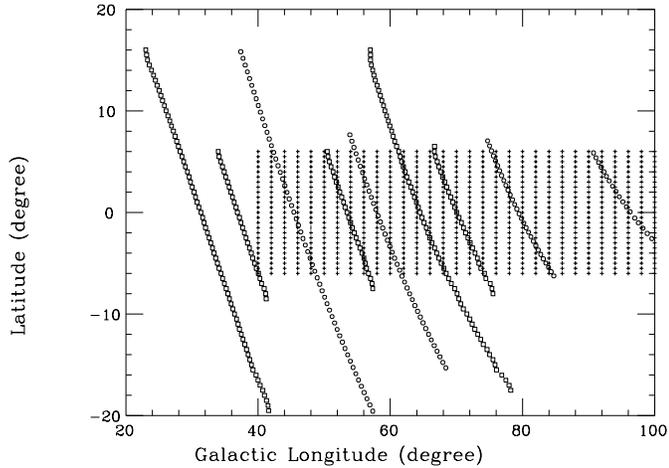}}
\caption{Zone of the galactic plane observed. We represent the 9 scans at
 constant declination and the mesh where we interpolate with constant
 $\triangle  l=2^\circ$ and constant $\triangle b=0.5^\circ$.}
\label{scan}
\end{figure}

In Fig.~\ref{scan} we show the scans used in the present work.
An interpolation technique was carried out in order to have
values at each point in a mesh (also reproduced in the figure)
with constant $\triangle l= 2^\circ$ and constant $\triangle b=0.5^\circ$

\section { Results}

To isolate the contribution from the Local Arm, it is
necessary to subtract the background emission, which is mainly
due to the exponential component of the disc. It is very
important to take into account the north warp, for which the
maximum deviation is  near the direction of interest here. Moreover,
both the warp and the optical Local Arm are observed over the
mean galactic plane. It is therefore necessary to use a model of
the exponential component of the warped disc.

In a recent paper, Porcel et al. (1997)
developed a model of the warped disc, which was used to
interpret the data from DIRBE (``Diffuse Infrared Background Experiment '')
(Freudenreich et al. 1994) on board COBE (``Cosmic Background Experiment'').
They have shown that either the stellar disc is noticeably less
warped than the gas, or a truncation of the stellar disc, taking
place not far  from the Sun, prevents  comparison between the
stellar and the gaseous warps. In this latter case, the
truncation radius cannot be larger than 13 kpc.

We will now use the same model to undertake the subtraction.
First of all, we must adjust the parameters of this model
 for a better fit to the COBE  observations. Suppose that we adopt
the first interpretation, i.e. that the stellar warp is smaller than
the gas one. Suppose that both warps are related at all
galactocentric radii by a constant. In other words, let $w_s(r)$
be the stellar warp curve and $w_g(r)$ the gaseous warp curve:
then, we assume $w_s(r)= \kappa \; w_g(r)$, where $\kappa$ is less
than unity and is one of our adjustable parameters.

The other two adjustable parameters are $z_\odot$, the Sun's height
over the mean plane and $\theta_{max}$, the galactocentric
azimuth of the maximum deviation from the mean plane due to the
warp. After comparison of the COBE results with a family of our
three-parametric models, we deduced that the best fit is
obtained for the following set of parameters: $z_\odot=10 \;
{\rm pc}$, $\theta_{max}=90^\circ$, $\kappa=0.6$. To obtain this family of parameters we benefited from the fact that they can be in principle obtained independently. The value of the mean $z$ depends on $z_\odot$ and not on the others parameters. The value of $\theta_{max}$ similarly is obtainable from a displacement of the $\theta$ origin alone, independent of the other parameters. These facts provided a very limited range of possible values, which  were, on the other hand, similar to current values typically adopted in other papers. What was considered a main object of this paper, and therefore with much more detail, was the obtention of the value of $\kappa$, considered implicitly to be one in most previous papers. A numerical least  squares method  provided the value of $\kappa=0.6$. After trials with different close values we estimated the error to be about 0.1.

The first two values for $z_\odot$ and $\theta_{max}$ are
typical,  just confirming those found by various authors, and lie
close to currently adopted values. The value of $\kappa$, on the
other hand, is striking, as unity is an implicitly assumed
value in most gravitational models of the warp  (see Binney
1992  and Combes 1994, for a review). Under this
interpretation, the stellar warp is roughly half the gas warp.

We have not included the extinction in our computations. Being the infrared extinction coefficient so small, it was demonstrated in the previous model by Porcel et al.(1997) that the results are noticeably insensitive to extinction, even when studying so distant features as warps. The effect of extinction is clearly much less important for a very close feature, as the Local Arm.

\begin{figure}
\resizebox{8.8cm}{!}{\includegraphics{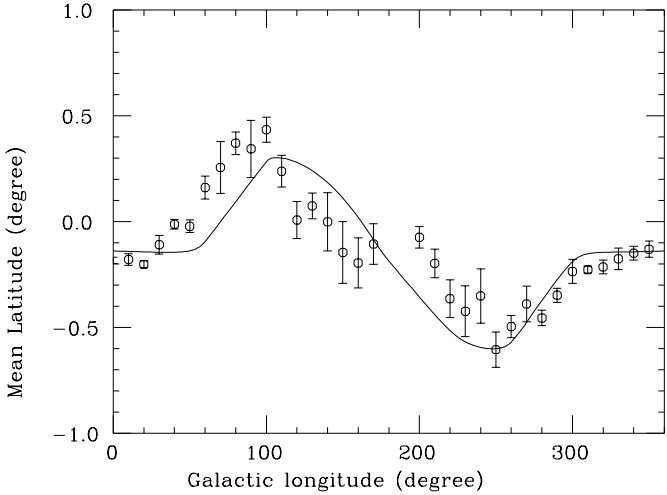}}
         \caption[]{Open circles stand for mean latitude of DIRBE data in the L band. The solid line shows the model's prediction with the parameters that best fit the observational data, i.e.  $z_\odot=10 \;
{\rm pc}$, $\theta_{max}=90^\circ$, $\kappa=0.6$ .\label{L-band}}
    \end{figure}

Within the region of the Local Arm, we now subtract  the data 
of the best fit model with $z_\odot=10 \;
{\rm pc}$,  $\theta_{max}=90^\circ$ and $\kappa=0.6$ from the
near-infrared data obtained by the ground-based telescope at
Tenerife  to obtain a
clear map of the Local Arm. The mean latitude of the L-band flux
predicted by the best fit model is reproduced in
Fig.~\ref{L-band}, together with the same data from  DIRBE.
The fitting cannot account for the observational values in
galactic latitudes $70^\circ-100^\circ$, i.e. the region of the
Local Arm.

\begin{figure}
\resizebox{8.8cm}{!}{\includegraphics{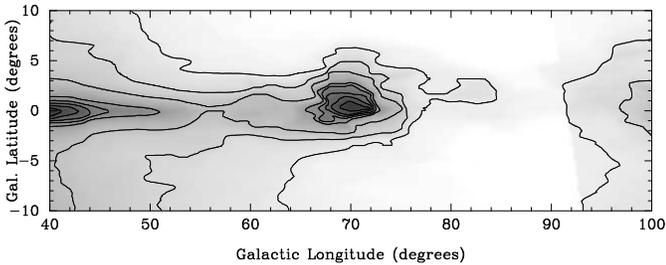}}
         \caption[]{Contour map of the brightness of the Galaxy 
DIRBE map in 12 $\mu $m in the longitude range  $40^\circ-100^\circ$.
 The contour levels are in MJy/sr from 17 to 33. The difference between  successive
 contour levels is 2 MJy/sr.\label{dirbe}}
    \end{figure}

Of course, the Local Arm is necessarily present in the DIRBE data.
This is clear in Fig.~\ref{dirbe}, where we show the
12 $\mu$m flux contour maps obtained from DIRBE data for the
region of interest.

Our procedure is correct because of the very different angular
extent of the warp and the Local Arm. We  used $360^\circ$
data to deduce the best fit model using DIRBE, and then 
applied this model to study a small feature, as the Local Arm
is only  $\approx 30^\circ$  in longitude.
 Fortunately for us, the Local Arm is larger (see for instance,
Becker \& Fenkart, 1970) but is angularly small because it passes
through the Sun. To obtain the fitting parameters the zone
outside the Local Arm was considered with a greater weight.

Suppose, on the other hand, that we adopt the second
interpretation of the apparent smallness of the stellar warp, i.e.
that it is due to a truncation of the stellar disc, which therefore does
not  reach the zone of highest warp. Though this
explanation is plausible and physically very different in the
interpretation of warps, the distinction is not important for
our present purposes, because the truncated disc produces  the
same map to be subtracted from the ground-based data.

\begin{figure}        
\resizebox{8.8cm}{!}{\includegraphics{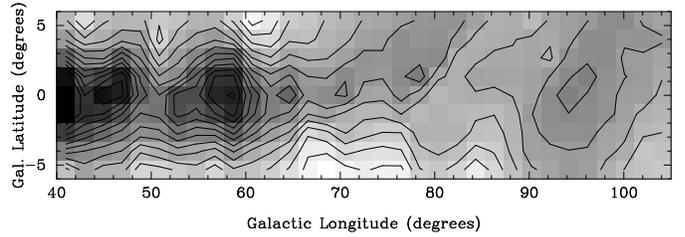}}
\caption[]{ Contour map of star counts up to 10 magnitude.
 The value between  successive contour levels is 100 $stars/degree^2$
 from 600 to 1800.\label{mapa10}}
   \end{figure}

In  Fig.~\ref{mapa10} we plot the contour maps corresponding to
the Tenerife observations, for the number of stars with $m \leq 10$.

\begin{figure}        
\resizebox{8.8cm}{!}{\includegraphics{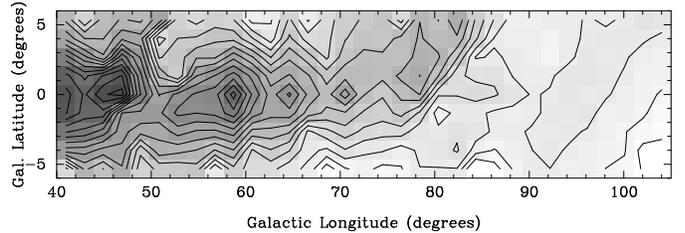}}
\caption[]{Contour map of star counts up to 8 magnitude.
 The value between  successive contour levels is 20 $stars/degree^2$
 from 100 to 440.
\label{mapa8}}
\end{figure}

\begin{figure*}        
\resizebox{18cm}{!}{\includegraphics{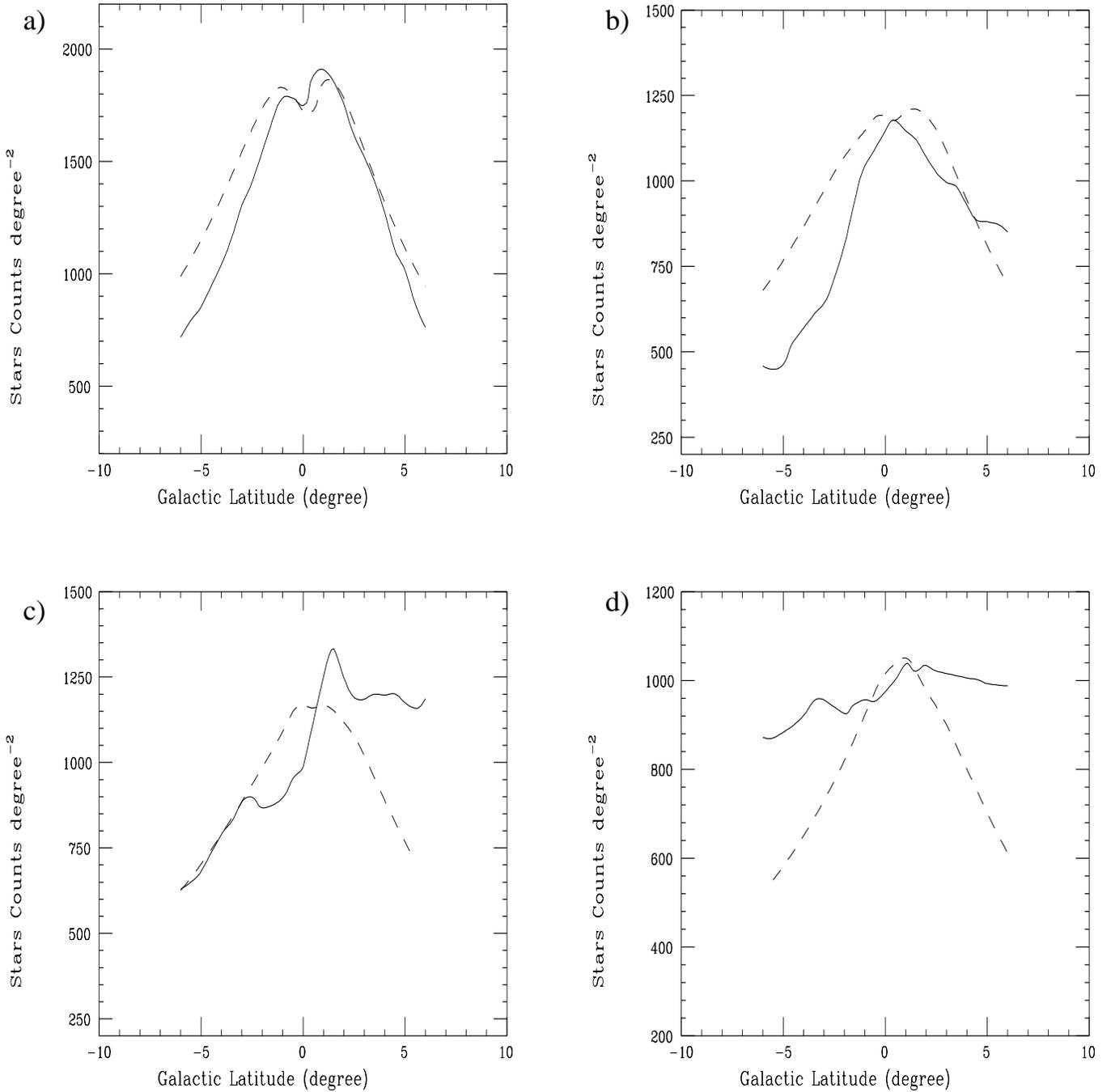}}
\caption[]{Latitude profile of the star counts up to 10 magnitude. Solid line shows the observational data and dashed line the model prediction:  $z_\odot=10 \;
{\rm pc}$, $\theta_{max}=90^\circ$, $\kappa=0.6$. (a) for $l=60^\circ$, (b) for $70^\circ$ (c) 
for $l=80^\circ$ (d) for $l=90^\circ$. \label{perfil}}
   \end{figure*}

Figure~\ref{mapa8}  shows the same plot for stars with $m \leq
8$. The first map includes nearly all observed stars; the second
map may be more interesting for our purposes, as nearly all $m \leq
8$ stars belong to the Local Arm. Some $m \leq 8$  bright stars may not belong to the Local Arm, but the number of $m \leq 8$  distant stars must be so scarce, that the distribution of $m \leq 8$ in the expected region of the Local Arm is the best way to define its geometry. Latitude profiles for four
longitudes, $ l=60^\circ$, $70^\circ$, $80^\circ$ and $90^\circ$
are very illustrative to see how the contribution of the Local
Arm is apparent over the emission of the rest of the  disc
(Figs. \ref{perfil}a-d). These profiles reasonably match what
is to be expected from the  model outside the Local Arm region. 
This fact is
appreciated at longitude $60^\circ$ where the arm lies outside,
but in the Local Arm  the profile varies from the model. At
$90^\circ$,  the great extension in latitude  is noticeable, as
observational data do not show any clear dependence on latitude.

\begin{figure}
\resizebox{8.8cm}{!}{\includegraphics{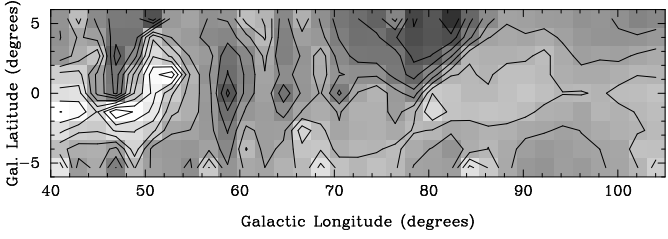}}
\caption[]{
Contour map of star counts up to 8 magnitude due to the Local Arm,
 after subtracting the contribution of the exponential component of
 the disc. The value between  successive contour levels is 20
 $stars/degree^2$ from -30 to 170.
 \label{brazo}}
   \end{figure}

Figure \ref{brazo} is the final objective of this work, which is
 clear contour maps of the Local Arm without contamination from
the exponential component of the warped thin disc. The Arm is
clearly seen with values of $m \leq 8$ stars per square
degree higher than 130, at a longitude of $80^\circ$ in the
North. There is also visible a large extension in latitude at longitudes
between $\approx 85^\circ$ and $90^\circ$. This thickening of
the disc is produced by the proximity of the Local Arm to the
Sun. We then see two properties of the Arm. It is oriented towards  the Sun
with a net elevation above the disc and is (angularly) very thick.

\section{Conclusions} 
\begin{figure}
\resizebox{8.8cm}{!}{\includegraphics{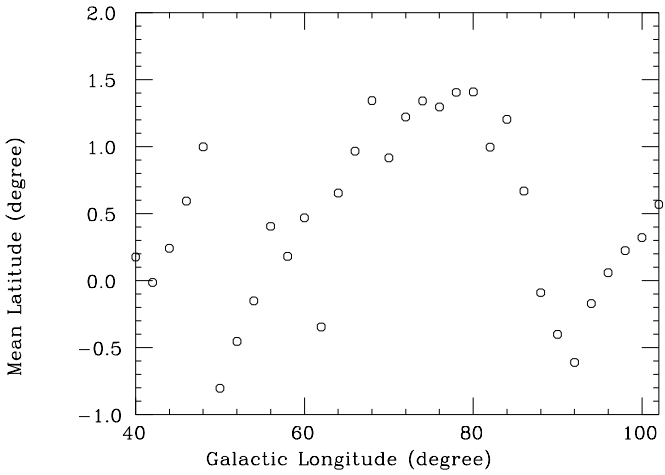}}
\caption[]{Mean latitude for the profile of star counts after
 subtracting the exponential component of the disc. The Local Arm begins
 at  l=$70^\circ$. \label{medio}}
     \end{figure}

In order to better interpret  Fig. \ref{brazo}, we
calculated the curve [mean b,l] to quantify the elevation of the
Local Arm over the mean plane. This is depicted in Fig.~\ref{medio}.

\begin{figure}
\resizebox{8.8cm}{!}{\includegraphics{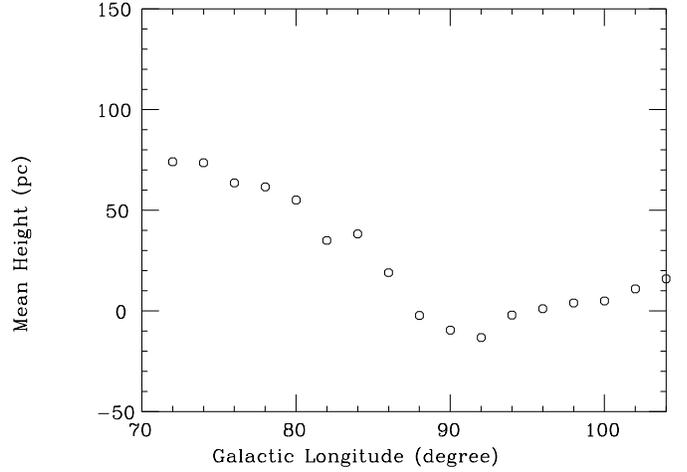}}
    \caption[]{Mean height of the Local Arm in galactic longitude. \label{zlongitud}}
\end{figure}

\begin{figure}
\resizebox{8.8cm}{!}{\includegraphics{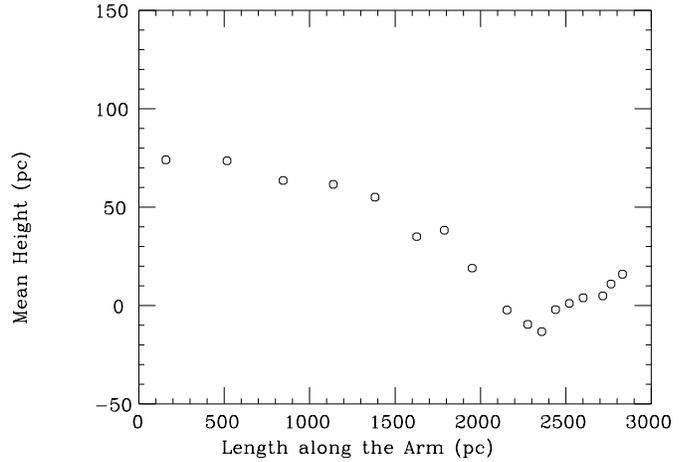}}
    \caption[]{Mean height of the Local Arm related to the real
 distance: the length along the arm. \label{zporbrazo}}
     \end{figure}

As the projected configuration of the arm is reasonably well known
from Becker \& Fenkart (1970) we have also plotted [mean z,l]
in Fig. \ref{zlongitud}. The projected configuration of the arm is a logarithmic spiral with parameters taken from Wainscoat et al. (1992). Furthermore, longitude can be translated into
real distance $S$, which is exactly  the length along the arc, taking as
the origin a point in the arm where $l=70^\circ$, about 3000 pc
away. Figure \ref{zporbrazo} shows the curve [mean z,$S$]. The arm
is elevated about 70 pc and this elevation decreases when the arm
comes closer to the Sun, where the elevation practically
vanishes. The other part of the arm lies in the direction around
$l=270^\circ$ and cannot be observed in the northern hemisphere.

\begin{figure}
\resizebox{8.8cm}{!}{\includegraphics{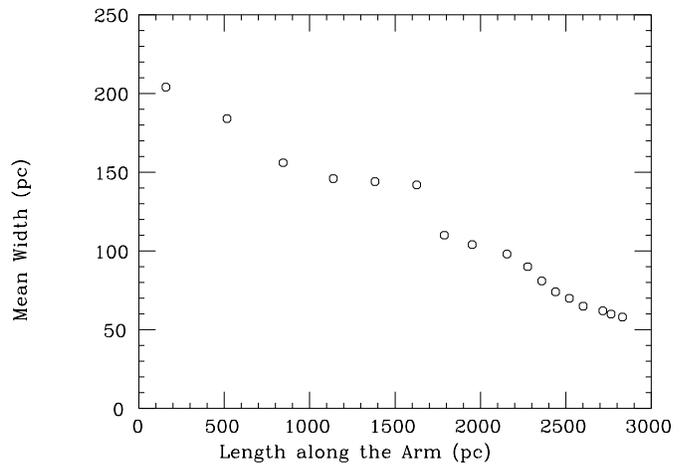}}
    \caption{Mean width of the Local Arm in relation to the distance 
along the arm.\label{ancho}}
\end{figure}

We have also calculated the width of the arm, and Fig. \ref{ancho}
shows the [width,$S$] curve. It must be interpreted with caution:
as  scans are limited in latitude ($|b| \leq 6^\circ$) the arm
width and the arm mean position are underestimated. We have seen that the angular width of
the arm becomes very large close to the Sun. But when angles are
translated to real widths, we see that the arm is thinner as the Sun
is approached. At $S$=0 the width is larger than 200 pc.

In  Fig. \ref{brazo} a sharp peak is observed  around $l=60^\circ$. The identification of this feature is not easy but it is clear that it corresponds to a real peak in the number of stars found in that direction, and not to an error in the reduction processes. Such a peak is not observed in the DIRBE data. Possible explanations should take into account the patchy nature of the absorbing clouds, and the different resolution of DIRBE and our data, but we have not found a relation with any other known feature of the galaxy, nor with detectable especial features in CO or 21 cm maps.

At a galactic latitude around $80^\circ$, a large and unexpected deviation of the arm with respect to the mean plane prevents us to obtain  reasonable values of the mean position of the Local Arm, so that the values in our Fig. 8 are highly underestimated. The Arm is to a great extent outside the small observation window of $\triangle b = \pm 6^\circ$. Clearly, future observations should be carried out for $|b| \leq 12 ^\circ$ in order to assure better observing conditions and better quantitative results. Nevertheless, even if incomplete, our map in figure 8 constitutes a first observation at this wavelength. This fact together with the substraction of the foreground sources which has been carried out, makes our observations an important source of information to study the closest known spiral arm.

There is a very noticeable agreement between our [mean z,l]
curve and the similar one obtained by  Kolesnick \& Vedenicheva (1978).
This agreement is mainly due to the fact that we are actually
observing young stars. Near infrared surveys are usually used to
trace the distribution of old stars, but due to the proximity of
the Local Arm, what we are mainly observing are OB stars with
$M_K \approx -4.7$. Nevertheless observations at different
wavelengths must be obtained even if they confirm the optical
description. Our map, on the other hand, is uncontaminated by
other sources, i.e. by the exponential component of the disc,
warp included. This is important, as the positions of the arm and
the north warp maximum are found in the same direction.

\begin{figure}
\resizebox{8.8cm}{!}{\includegraphics{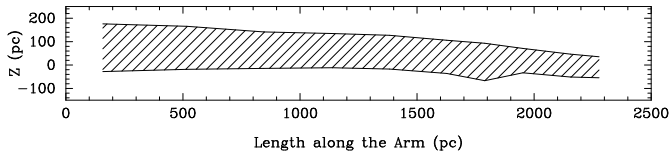}}  
\caption{Geometrical configuration of
the Local Arm viewed from the galactic centre. \label{forma}}
\end{figure}

Figure \ref{forma} represents the geometrical configuration of
the Local Arm, when  viewed from approximately the galactic centre.

\begin{acknowledgements}
The COBE data sets were developed by the NASA Goddard Space Flight Center
 under the guidance of the COBE Science Working Group and were provided
 by the NSSDC.
\end{acknowledgements}

\end{document}